\documentstyle[preprint,amssymb,aps]{revtex}
\tightenlines
\begin{document}
\preprint{ch3f.tex}
\draft
\title{On spin-rotation contribution to nuclear spin conversion in\\
$\bbox{{\rm C}_{3{\rm v}}}$-symmetry molecules. Application to $%
\bbox{{\rm
CH}_3{\rm F}}$}
\author{K. I. Gus'kov\thanks{%
Electronic address: fractal@iae.nsk.su}}
\address{Institute of Automation and Electrometry, \\
Siberian Branch of the Russian Academy of Sciences,\\
Novosibirsk, 630090, Russia}
\date{\today }
\maketitle

\begin{abstract}
The  symmetrized  contribution    of    $E$-type    spin-rotation
interaction to conversion between spin modifications of $E$-  and
$A_1$-types in molecules with ${\rm  C}_{3{\rm  v}}$-symmetry  is
considered.    Using  the  high-$J$  descending  of   collisional
broadening for accidental  rotational  resonances  between  these
spin  modifications,    it  was  possible  to  co-ordinate    the
theoretical  description  of  the  conversion   with    (updated)
experimental  data  for  two  carbon-substituted   isotopes    of
fluoromethane. As a result, both $E$-type spin-rotation constants
are obtained. They are roughly one and a half times more than the
corresponding constants for (deutero)methane.

\end{abstract}

\pacs{PACS numbers: 31.30.Gs, 33.50.-j}


\section{Introduction}

The purpose of the present work is to compare the theoretical description of
spin-modification (or, differently, spin-isomer) conversion \cite
{Cha91,Gus95} with updated experimental data for two carbon-substituted
isotopes of fluoromethane \cite{NBHC98,NSCH95,NSCH96}. The main difficulties
on this way are conditioned, firstly, by the lack of data about $E$-type%
\footnote{%
See Eq.~(\ref{Vsr:sym:double}) below.} spin-rotation constants of this
molecule and, secondly, by the insufficiently reliable data about values and
collisional broadenings of conversional accidental resonances of rotational
levels of its spin modifications. Resigning oneself to the latter, we
attempt to extract at least preliminary information on this constants by
reversing the problem on the conversional contribution of spin-rotation
interaction. It is useful to compare the obtained constants for
fluoromethane with the known ones for methane presenting them in the
similarly orientated molecule-fixed frames.

In the beginning, it is necessary to make two corrections.\footnote{%
My attention to them was turned by P. L. Chapovsky. For comparison, see also
\cite{IB98}.} The consideration of spin-modification conversion for ${\rm C}%
_{3{\rm v}}$-symmetry molecules in \cite{Gus95} explores the spin-rotation
Hamiltonian form of limited applicability. Being widely encountered in
papers and books (see, e.g., \cite{TS55,Fly78}), this form is non-Hermitian
for nonspherical tops of nonlinear molecules. It is necessary to use the
standard symmetrizing procedure \cite{Mes61} in order to come to Hermitian
form of the Hamiltonian. The second correction deals with the contribution
of Thomas precession to the spin-rotation constants \cite{GMTVV54}. As the
most important consequence, the components of these tensorial constants even
in $R$-forms (\ref{Rform}) may have asymmetry, i.e., transposed matrices $%
\pmb{\sf R}^{\varsigma \top }\neq \pmb{\sf R}^\varsigma $ for off-axial
nuclei. The number of independent components of $R$-form (\ref{RH1}) may
grow up to five, i.e., two of $A_1$-type and three (but not two as in \cite
{Gus95}) of $E$-type.

The two carbon-substituted isotopes of fluoromethane (${}^{12,13}{\rm CH}_3%
{\rm F}$) in the room-\-temp\-e\-ra\-ture gaseous phase have essentially
different rates of mutual conversion of para- and ortho-modification.
According to \cite{NBHC98,NSCH95,NSCH96,Nag98}, the updated
pressure-normalized rates of spin-modification conversions, i.e. $\gamma
^{(P)}=\gamma /P$, are\footnote{%
In this paper all the rates (or frequencies) are angular but measured in $%
{\rm Hz}=2\pi ~{\rm s}^{-1}$.}
\begin{equation}
{}^{12}\gamma ^{(P)}=35.3(31)~\mu {\rm Hz}/{\rm Torr}  \label{12exp}
\end{equation}
and
\begin{equation}
{}^{13}\gamma ^{(P)}=1942(95)~\mu {\rm Hz}/{\rm Torr}.  \label{13exp}
\end{equation}
It may be compared with the $P$-normalized collisional broadening
(halfwidth) of rotational lines
\[
\Gamma _{J^{\prime }K^{\prime }/JK}^{(P)}=\Gamma _{J^{\prime }K^{\prime
}/JK}/P
\]
where
\begin{equation}
\Gamma _{J^{\prime }K^{\prime }/JK}=(\Gamma _{J^{\prime }K^{\prime }}+\Gamma
_{JK})/2.  \label{Gamma}
\end{equation}
According to \cite{JPN:73}, we have
\[
\Gamma _{5,3/4,3}^{(P)}=15.2(8)~{\rm MHz}/{\rm Torr}.
\]
The theoretical estimates of nuclear spin-spin contribution to $\gamma
^{(P)} $ were obtained in \cite{Cha91} assuming the same broadening for all
the conversional transitions. These estimates make up $7\%$ of (\ref{12exp})
for ${}^{12}{\rm CH}_3{\rm F}$ and $43\%$ of (\ref{13exp}) for ${}^{13}{\rm %
CH}_3{\rm F}$. But this simple assumption leads to an incompatibility
between (\ref{12exp}) and (\ref{13exp}), if, following \cite{Gus95}, one
attempts to add the spin-rotation contribution to the spin-spin one. The
rejection of the last high-$J$ para-ortho resonance standing out against the
rest for ${}^{12}{\rm CH}_3{\rm F}$ might be a possible way out of this
situation. The reason for that has been found in \cite{TEB79}. It turned out
that the collisional broadening of rotational lines is decreased with
increasing $J$. As a result, the contribution of the high-$J$ resonance is
effectively suppressed and the indicated incompatibility is eliminated.

It make sense to collect together all spin-rotation constants of
fluoromethane and (deutero)methane. Further, comparing them, we can make the
definite conclusion on signs of the $E$-type constants for fluoromethane.

\section{Symmetrized form of spin-rotation Hamiltonian}

The spin-rotation interaction appears as a consequence of action of a local
magnetic field $\bbox{\hat{B}}^\varsigma $ on the magnetic moment $%
\bbox{\hat{\mu}}_I^\varsigma =\mu _I^\varsigma \bbox{\hat{I}}^\varsigma $ of
nucleus labelled by $\varsigma $. Here $\mu _I^\varsigma =g_I^\varsigma \mu
_{{\rm N}}$, $g_I^\varsigma $ is the nuclear $g$-factor, $\mu _{{\rm N}}$ is
the nuclear magneton, and $\bbox{\hat{I}}^\varsigma $ is the nuclear spin
undimensioned by Planck's constant $\hbar $. The intramolecular field $%
\bbox{\hat{B}}^\varsigma $ is conditioned by rotation of nuclear framework
and internal motions of electrons. The non-symmetrized Hamiltonian of this
interaction is
\begin{equation}
\hat{V}_{\text{sr}}=-\sum_\varsigma \bbox{\hat{B}}^\varsigma \cdot %
\bbox{\hat{\mu}}_I^\varsigma  \label{Vsr}
\end{equation}
where $\bbox{\hat{B}}^\varsigma =\bbox{\hat{J}}\cdot \pmb{\sf C}^\varsigma
/\mu _I^\varsigma $, $\bbox{\hat{J}}$ is rotational angular momentum (also
undimensioned by $\hbar $), and $\pmb{\sf C}^\varsigma $ is spin-rotation
tensor. The energy (as well as $\pmb{\sf C}^\varsigma $) is measured in
frequency units. The convenient unit of measurement for $\pmb{\sf C}%
^\varsigma $ usually is ${\rm kHz}$. In molecule-fix frame\footnote{%
Projections onto this frame are underscored. E.g., Cartesian ones $\hat{J}_{%
\underline{i}}=\bbox{u}_{\underline{i}}\cdot \bbox{\hat{J}}=\bbox{\hat{J}}%
\cdot \bbox{u}_{\underline{i}}$.} the components of $\underline{\pmb{\sf C}}%
^\varsigma $ are constant but $[\hat{J}_{\underline{i}},\hat{I}_{\underline{j%
}}^\varsigma ]=-i\varepsilon _{\underline{i}\underline{j}\underline{k}}\hat{I%
}_{\underline{k}}^\varsigma $ unlike $[\hat{J}_i,\hat{I}_j^\varsigma ]=0$ in
space-fix frame. Hence, by Hermitian conjugation, we have operator $\hat{V}_{%
\text{sr}}^{\dagger }\neq \hat{V}_{\text{sr}}$. To convert $\hat{V}_{\text{sr%
}}$ into Hermitian one, we must use the standard symmetrization procedure
\cite{Mes61} (see also, as an example, \cite{Loe83}), i.e., $\hat{J}_{%
\underline{i}}\hat{I}_{\underline{j}}^\varsigma \rightarrow (\hat{J}_{%
\underline{i}}\hat{I}_{\underline{j}}^\varsigma +\hat{I}_{\underline{j}%
}^\varsigma \hat{J}_{\underline{i}})/2$. As a result, we come to symmetrized
spin-rotation Hamiltonian
\begin{equation}
\hat{V}_{\text{sr}}^{(+)}=(\hat{V}_{\text{sr}}+\hat{V}_{\text{sr}}^{\dagger
})/2=-\frac 12\sum_\varsigma (\bbox{\hat{J}}\cdot \pmb{\sf C}^\varsigma
\cdot \bbox{\hat{I}}^\varsigma +\bbox{\hat{I}}^\varsigma \cdot \pmb{\sf C}%
^{\varsigma \top }\cdot \bbox{\hat{J}}).  \label{Vsr:sym}
\end{equation}

The initial components $C_{ij}^\varsigma $ have reducible lower and upper
indices respectively for the group of arbitrary molecular rotations and
inversions, ${\rm O}_3={\rm SO}_3\otimes {\rm C}_{{\rm i}}$, and the group
of permutations of spatial and spin coordinates for identical nuclei, ${\rm S%
}_3\cong {\rm C}_{\underline{3{\rm v}}}\subset {\rm O}_{\underline{3}}$.
Transforming this components into double irreducible ones $C_{(11)\varkappa
_\varrho \dot{q}}^{\lambda \dot{s}}$ defined in \cite{Gus95}, we can also
obtain
\begin{equation}
\hat{V}_{\text{sr}}^{(+)}=-\frac 12\sum_{\varkappa \lambda }\sqrt{[\varkappa
][\lambda ]}\left\{ (-1)^\varkappa \left[ \bbox{\hat{J}}_1^{A_1}{}_{\otimes
}^{\otimes }\pmb{\sf C}_{(11)\varkappa }^\lambda {}_{\otimes }^{\otimes }%
\bbox{\hat{I}}_1^\lambda \right] _{0_{+}}^{A_1}+\left[ \bbox{\hat{I}}%
_1^\lambda {}_{\otimes }^{\otimes }\pmb{\sf C}_{(11)\varkappa }^\lambda
{}_{\otimes }^{\otimes }\bbox{\hat{J}}_1^{A_1}\right] _{0_{+}}^{A_1}\right\}
\label{Vsr:sym:double}
\end{equation}
where $\varkappa =0,1,2$ and $\lambda =A_1,E$ (the last type exists only for
off-axial nuclei). The single symbols in square brackets designate dimension
of their respective representations. E.g., in the case of the ${\rm SO}_3$
group, we have $[\varkappa ]=2\varkappa +1$. Just this form of spin-rotation
Hamiltonian is used in analitical calculations (cp. with \cite{Gus95}). It
is necessary to note directly now, that the non-commutation of $\hat{J}_{%
\underline{i}}$ and $\hat{I}_{\underline{j}}^\varsigma $ has a little
significance at high $J$. We shall see this situation does take place for
fluoromethane.

\section{Isotopically independent spin-rotation constants}

In order that the spin-rotation constants do not change with isotopical
substitutions, it is convenient to use their modified $R$-form, i.e., let
\begin{equation}
\pmb{\sf C}^\varsigma =g_I^\varsigma \tilde{\pmb{\sf B}}\cdot \pmb{\sf R}%
^\varsigma .  \label{Rform}
\end{equation}
Here $\tilde{\pmb{\sf B}}=2\pmb{\sf B}=\hbar \pmb{\sf I}^{-1}$, i.e.
proportional to the inverse tensor of inertia moment localized (as well as $%
\bbox{\hat{J}}$) in inertia center. The factor $\hbar $ is chosen for $%
\tilde{\pmb{\sf B}}$ to have frequency dimensions as well as $\pmb{\sf C}%
^\varsigma $. The tensor $\pmb{\sf B}$ is defined by rotational spectrum of
molecule (\ref{six}). Choosing the molecule-fix frame with its $z $-axis
along the axis of molecular symmetry (${\rm C}\rightarrow {\rm F}$) we
obtain the diagonal
\begin{equation}
\underline{\tilde{\pmb{\sf B}}}=
\begin{array}[b]{cc}
&
\begin{array}{cc}
\stackrel{\underline{\bot }}{\hphantom{\tilde{B}_{\underline{\bot }}
\hat{\pmb{\sf 1}}_2}} & \stackrel{\underline{z}}{\hphantom{\tilde{B}_{%
\underline{z}}}}
\end{array}
\\
\begin{array}{c}
{\scriptstyle\underline{\bot }} \\
{\scriptstyle\underline{z}}
\end{array}
& \left[
\begin{array}{cc}
\tilde{B}_{\underline{\bot }}\hat{\pmb{\sf 1}}_2 & \bbox{0} \\
\bbox{0}^{\top } & \tilde{B}_{\underline{z}}
\end{array}
\right] .
\end{array}
\label{tildeB}
\end{equation}
For isotopes ${}^{12,13}{\rm CH}_3{\rm F}$ we have transverse components $%
{}^{12}\tilde{B}_{\underline{\bot }}=51.1(1)~{\rm GHz}$ and ${}^{13}\tilde{B}%
_{\underline{\bot }}=49.7(1)~{\rm GHz}$, both longitudinal ones ${}^{12,13}%
\tilde{B}_{\underline{z}}=310.7(1)~{\rm GHz}$. The non-dimensional tensor $%
\pmb{\sf R}^\varsigma $ is only defined by electric charge distribution. It
is one and the same for all the isotopes of molecule in their invariable
geometry. Another attractive property of $\pmb{\sf R}^\varsigma $ is much
greater symmetry in comparison with $\pmb{\sf C}^\varsigma $, i.e., we may
have $\pmb{\sf R}_{(-)}^\varsigma =0$ but $\pmb{\sf C}_{(-)}^\varsigma \neq 0
$ where $\pmb{\sf R}_{(\pm )}^\varsigma \equiv (\pmb{\sf R}^\varsigma \pm %
\pmb{\sf R}^{\varsigma \top })/2$. At last, it is known \cite{GF72} that
\begin{equation}
\pmb{\sf R}^\varsigma \simeq m_{{\rm e}}\bbox{\sigma}_{\text{ma}}^\varsigma
/m_{{\rm p}}  \label{shielding}
\end{equation}
where tensor $\bbox\sigma _{\text{ma}}^\varsigma =\bbox\sigma _{\text{%
molecule}}^\varsigma -\bbox\sigma _{\text{atom}}^\varsigma $, i.e. the
difference of nuclear magnetic shielding for nucleus $\varsigma $ by forming
the molecule from free atoms. Thus, it is convenient to measure $R$%
-constants in ${\rm ppb}=10^{-9}$, i.e. parts per billion, as well as $%
\sigma $-constants in ${\rm ppm}=10^{-6}$.

Some (most likely small) asymmetry of tensor $\pmb{\sf R}^\varsigma $ may be
for off-axial spins by taking into account the Thomas correction dependent
on motion of spinning nuclei \cite{GMTVV54}. We are going to neglect this
asymmetry for fluoromethane, having in view of the zero asymmetric part and
small contribution of the Thomas correction in the symmetric one \cite{AR66}
for methane used here as a sample.

Now let us devide, as in \cite{Gus95}, all the nuclei $\varsigma$ into axial
(vertical) $v={\rm F,C}$ and off-axial (horizontal) $h={\rm H}^1, {\rm H}^2,
{\rm H}^3$. The use of the C$_{\underline{3{\rm v}}}$-symmetry group for
off-axial nucleus ${\rm H}^1$ leads to the following form of tensor
\begin{eqnarray}
\underline{\pmb{\sf R}}^{{\rm H}^1} &=&
\begin{array}{cc}
&
\begin{array}{ccc}
\stackrel{\underline{x}}{\hphantom{R^{{\rm H}E}_{(+) \underline{z}}-R^{{\rm
H}E}_{(-) \underline{z}}}} & \stackrel{\underline{y}}{\hphantom{R^{{\rm
H}A_1}_{\underline{ \bot }}-R^{{\rm H}E}_{\underline{\bot }}}} & \stackrel{%
\underline{z}}{\hphantom{R^{{\rm H}E}_{(+)\underline{z}} +R^{{\rm
H}E}_{(-)\underline{z}}}}
\end{array}
\\
\begin{array}{c}
{\scriptstyle\underline{x}} \\
{\scriptstyle\underline{y}} \\
{\scriptstyle\underline{z}}
\end{array}
& \left[
\begin{array}{ccc}
R_{\underline{\bot }}^{{\rm H}A_1}+R_{\underline{\bot }}^{{\rm H}E} & 0 &
R_{(+)\underline{z}}^{{\rm H}E}+R_{(-)\underline{z}}^{{\rm H}E} \\
0 & R_{\underline{\bot }}^{{\rm H}A_1}-R_{\underline{\bot }}^{{\rm H}E} & 0
\\
R_{(+)\underline{z}}^{{\rm H}E}-R_{(-)\underline{z}}^{{\rm H}E} & 0 & R_{%
\underline{z}}^{{\rm H}A_1}
\end{array}
\right]
\end{array}
\nonumber \\
&=&
\begin{array}[b]{cc}
&
\begin{array}{cc}
\stackrel{\underline{\bot }}{\hphantom{(R^{{\rm H}E}_{(+)
\underline{z}}-R^{{\rm H}E}_{(-) \underline{z}})\bbox{n}^{1\top}}} &
\stackrel{\underline{z}}{\hphantom{(R^{{\rm H}E}_{(+)\underline{z}} +R^{{\rm
H}E}_{(-)\underline{z}})\bbox{n}^1}}
\end{array}
\\
\begin{array}{c}
{\scriptstyle\underline{\bot }} \\
{\scriptstyle\underline{z}}
\end{array}
& \left[
\begin{array}{cc}
R_{\underline{\bot }}^{{\rm H}A_1} \hat{\pmb{\sf 1}}_2+R_{\underline{\bot }%
}^{{\rm H}E}\bbox{\hat{\sigma}}_{{\rm v}}^{{\rm H}^1} & (R_{(+)\underline{z}%
}^{{\rm H}E}+R_{(-)\underline{z}}^{{\rm H}E})\bbox{n}^{{\rm H}^1} \\
(R_{(+)\underline{z}}^{{\rm H}E}-R_{(-)\underline{z}}^{{\rm H}E})\bbox{n}^{%
{\rm H}^1\top } & R_{\underline{z}}^{{\rm H}A_1}
\end{array}
\right]
\end{array}
.  \label{RH1}
\end{eqnarray}
In other cases $\bbox{\hat{\sigma}}_{{\rm v}}^{{\rm H}^1}\rightarrow %
\bbox{\hat{\sigma}}_{{\rm v}}^h$ and $\bbox{n}^{{\rm H}^1}\rightarrow %
\bbox{n}^h$ as describing in \cite{Gus95}. The same symmetry of
spin-rotation interaction for axial nuclei $v$ retains only $A_1$-type
components, i.e.,
\begin{equation}
\underline{\pmb{\sf R}}^v=\left[
\begin{array}{cc}
R_{\underline{\bot }}^{vA_1}\hat{\pmb{\sf 1}}_2 & \bbox{0} \\
\bbox{0}^{\top } & R_{\underline{z}}^{vA_1}
\end{array}
\right] .  \label{Rv}
\end{equation}
If we designate the components of $\underline{\pmb{\sf C}}^{{\rm H}^1}$ and $%
\underline{\pmb{\sf C}}^v$ by analogy with (\ref{RH1}) and (\ref{Rv}) then
\begin{eqnarray}
&\displaystyle C_{\underline{\bot }}^{\varsigma A_1}=g_I^\varsigma \tilde{B}%
_{\underline{\bot }}R_{\underline{\bot }}^{\varsigma A_1},\quad C_{%
\underline{z}}^{\varsigma A_1}=g_I^\varsigma \tilde{B}_{\underline{z}}R_{%
\underline{z}}^{\varsigma A_1},  \nonumber \\
&\displaystyle C_{\underline{\bot }}^{{\rm H}E}=g_I^{{\rm H}}\tilde{B}_{%
\underline{\bot }}R_{\underline{\bot }}^{{\rm H}E},\quad C_{(\pm )\underline{%
z}}^{{\rm H}E}=g_I^{{\rm H}}\left( \tilde{B}_{(\pm )}R_{(+)\underline{z}}^{%
{\rm H}E}+\tilde{B}_{(\mp )}R_{(-)\underline{z}}^{{\rm H}E}\right)
\label{CfromR}
\end{eqnarray}
where $\tilde{B}_{(\pm )}=(\tilde{B}_{\underline{\bot }}\pm \tilde{B}_{%
\underline{z} })/2$. Here $\tilde{B}_{(-)}$ and $C^{{\rm H}E}_{(-)
\underline{z}}$ are defined with the reverse signs to \cite{Gus95}.

\section{The complete hyperfine contribution to spin-modification conversion
in $\bbox{{\rm C}_{3{\rm v}}}$-symmetry molecules}

The states of $E$- and $A_1$-types for three hydrogen spins of fluoromethane
differ in total spin $\bbox{\hat{I}}^{{\rm H}} =\sum_h\bbox{\hat{I}}^h$,
i.e., $I^{{\rm H}}=1/2$ (para) and $3/2$ (ortho) respectively. Under
equilibrium conditions, according to detailed balancing principle, the
increment and decrement of the populations of this states (in- and
out-terms) compensate each other, i.e., $\gamma _{\frac 12}\rho _{\frac 12%
}(\infty )=\gamma _{\frac 32}\rho _{\frac 32}(\infty )$. As a result, the
equilibrium ratio of the partial population to the total one is $\rho _{I^{%
{\rm H}}}(\infty )/\rho =(\gamma -\gamma _{I^{{\rm H}}})/\gamma $ ($\simeq
1/2$ for fluoromethane). Using light-induced drift method \cite{KPSC84}, we
can breakdown the equilibrium. Owing to the mutual conversion, the
non-equilibrium populations decay. This decay is described with one
exponent, i.e., $\delta \rho _{I^{{\rm H}}}(t)=\delta \rho _{I^{{\rm H}}}(0)%
{\rm e}^{-\gamma t}$ and $\sum_{I^{{\rm H}}}\delta \rho _{I^{{\rm H}}}(t)=0$%
. Let us consider now the expression for the conversional rate \cite{Cha91}
\begin{eqnarray}
\gamma &=&\gamma _{\frac 12}+\gamma _{\frac 32}  \nonumber \\
&=&\sum_{{}_{J\,K_0}^{J^{\prime }K_{r\neq 0}^{\prime }}}\sum_p\frac{2\Gamma
_{J^{\prime }K^{\prime }/JK}[{\cal W}_{\text{B}}^{(\frac 12)}(J_p^{\prime
}K_r^{\prime })+{\cal W}_{\text{B}}^{(\frac 32)}(J_pK_0)]}{\Gamma
_{J^{\prime }K^{\prime }/JK}^2+\omega _{J_p^{\prime }K_r^{\prime }/J_pK_0}^2}
\nonumber \\
&&\times \sum_{(a^{\prime })(a)}|\langle (a^{\prime })J_p^{\prime }K^{\prime
}|(\hat{V}_{\text{sr}}^{(+)}+\hat{V}_{\text{ss}})|(a)J_pK_0\rangle |^2.
\label{gamma}
\end{eqnarray}
We have described the symmetrized spin-rotation interaction $\hat{V}_{\text{%
sr}}^{(+)}$ in (\ref{Vsr:sym}) and (\ref{Vsr:sym:double}), the spin-spin one
$\hat{V}_{\text{ss}}$ in \cite{Gus95}. The halfwidth $\Gamma _{J^{\prime
}K^{\prime }/JK}$ was defined in (\ref{Gamma}). The frequency difference
\begin{equation}
\omega _{J_p^{\prime }K_r^{\prime }/J_pK_0}\equiv \omega _{J_p^{\prime
}K_r^{\prime }}-\omega _{J_pK_0}.
\end{equation}
Here $J\geq K_r\geq 0$ and $K$ is congruous with $r$ modulo $3$, i.e., $%
K\equiv r\pmod{3}$ where $r\in (0,1,2)$. The level parity
\begin{equation}
p=(-1)^{J+1}\delta _{K,0}\pm (1-\delta _{K,0}).  \label{parity}
\end{equation}
The rotational spectrum $\omega _{J_pK}$ may be expressed by the formula
\begin{eqnarray}
\omega _{J_pK} &=&B_KK^2+B_J\bbox{\hat{J}}^2-D_KK^4-D_{KJ}K^2 \bbox{\hat{J}}%
^2-D_J\bbox{\hat{J}}^4  \nonumber \\
&&+F_KK^6+F_{KJ}K^4\bbox{\hat{J}}^2 +F_{JK}K^2\bbox{\hat{J}}^4+F_J%
\bbox{\hat{J}}^6  \nonumber \\
&&+\delta _{K,3}F_J^{(3)}p(-1)^J\bbox{\hat{J}}^2 [\bbox{\hat{J}}^2-2] [%
\bbox{\hat{J}}^2-6].  \label{six}
\end{eqnarray}
Here, for short, $\bbox{\hat{J}}^2\equiv J(J+1)$. $B_K=B_{\underline{z}}-B_{%
\underline{\bot }}$ and $B_J=B_{\underline{\bot }}$. We take into account
quartic and sextic centrifugal distortions in the ground electronic and
vibrational states \cite{DBC:94}. Here the alphabetical choice for
designation of the coefficients is connected with the powers of
perturbations. Almost all the $BDF$-coefficients of isotopes $^{12,13}{\rm CH%
}_3{\rm F}$ may be found in \cite{PHC:93,PDW:94}. We assume the unknown $%
^{12}F_K$ is just a little less then the known $^{13}F_K$, i.e., $%
^{13}F_K-{}^{12}F_K\simeq 3~{\rm Hz}$.\footnote{%
Cp. with the less justified assumption ($^{12}F_K=0$) in \cite{Nag98}.\label
{FK}} $F_J^{(3)}$ defines $K$-doubling in the parity of rotational levels
with $K=3$. Its estimate can be found in \cite{CCDR93}.

The wave functions of every spin-rotation subsystem are $\left| (K_r)_{(J_p{%
I^{{\rm H}}})M_JM_I^{{\rm H}}}^{(\Gamma _J\Gamma _{I^{{\rm H}%
}})A_20}\right\rangle \prod_v|I^vM_I^v\rangle $ where the first factor is $%
\left| (K_{r\neq 0})_{(J_p\frac 12)M_JM_I^{{\rm H}}}^{(EE)A_20}\right\rangle
$ or $\left| (K_0)_{(J_p\frac 32)M_JM_I^{{\rm H}}}^{(A_2A_1)A_20}\right%
\rangle $ (see \cite{Gus95}). The populations of levels are characterized by
their Boltzmann factors
\[
{\cal W}_{\text{B}}^{({I^{{\rm H}}})}(J_pK_r)=\left. {\rm e}^{-\hbar \omega
_{J_pK_r}/k_{\text{B}}T}\right/ Q_{\text{sr}}^{({I^{{\rm H}}})}.
\]
The spin-rotation statistical sum
\[
Q_{\text{sr}}^{({I^{{\rm H}}})}=\prod_v[I^v]\,[{I^{{\rm H}}}%
]\sum_J[J]\sum_{K_rp}{\rm e}^{-\hbar \omega _{J_pK_r}/k_{\text{B}}T}.
\]
The square brackets, as defined after Eq.~(\ref{Vsr:sym:double}), designate
the degree of space-degeneracy (or statistical weight) for rotational
angular momentum or nuclear spins. By room temperature ($20~{}^o{\rm C}$) we
have $B_{\underline{\bot }}\ll k_{\text{B}}T/\hbar \simeq 6108.2~{\rm GHz}$
and
\[
Q_{\text{sr}}^{(\frac 12)}\simeq Q_{\text{sr}}^{(\frac 32)}\simeq
\prod_v[I^v]\,\frac{4k_{\text{B}}T}{3\hbar B_{\underline{\bot }}}\sqrt{\frac{%
\pi k_{\text{B}}T}{\hbar B_{\underline{z}}}}.
\]
Hence, for isotopes of fluoromethane, it follows
\[
{}^{12}Q_{\text{sr}}^{({I^{{\rm H}}})}\simeq \prod_v[I^v]\,3554,\quad
^{13}Q_{\text{sr}}^{({I^{{\rm H}}})}\simeq \prod_v[I^v]\,3625.
\]
Here it is convenient to represent already symmetrized form of expression
(92) from \cite{Gus95}:
\begin{eqnarray}
&&\sum_{(a^{\prime })(a)}|\langle (a^{\prime })J_p^{\prime }K_r^{\prime }|(%
\hat{V}_{\text{sr}}^{(+)}+\hat{V}_{\text{ss}})|(a)J_pK_0\rangle
|^2=\prod_{v^{\prime }}[I^{v^{\prime }}](1+\delta _{K^{\prime },0})(1+\delta
_{K,0})[J]  \nonumber \\
&&\times \left\{ \vphantom{ \sum_v }\frac{\delta _{|\Delta J|,1}}4\left[ %
\vphantom{ \left[\sqrt{(J')}\right]^2}\delta _{|\Delta K|,2}(C_{\underline{%
\bot }}^{{\rm H}E})^2\langle J^{\prime }K^{\prime }|2\Delta K~JK\rangle
^2(J^{\prime }+J+3)(J^{\prime }+J-1)\right. \right.  \nonumber \\
&&+\delta _{|\Delta K|,1}\left[ C_{(+)\underline{z}}^{{\rm H}E}\langle
J^{\prime }K^{\prime }|2\Delta K~JK\rangle \sqrt{(J^{\prime }+J+3)(J^{\prime
}+J-1)}\right.  \nonumber \\
&&\left. \left. \vphantom{\sqrt{(J')}}+C_{(-)\underline{z}}^{{\rm H}%
E}\langle J^{\prime }K^{\prime }|1\Delta K~JK\rangle \Delta K\Delta
J(J^{\prime }+J+1)\right] ^2\right] +\langle J^{\prime }K^{\prime }|2\Delta
K~JK\rangle ^2  \nonumber \\
&&\left. \times \left[ \delta _{|\Delta K|,2}\frac{27}{16}S_{{\rm HH}%
}^2+6\sum_v\bbox{\hat{I}}^{v2}\left( \delta _{|\Delta K|,2}\frac{r_{{\rm H}%
O}^2}4+\delta _{|\Delta K|,1}r_{vO}^2\right) \frac{r_{{\rm H}O}^2}{r_{{\rm H}%
v}^4}S_{{\rm H}v}^2\right] \right\} .  \label{conv}
\end{eqnarray}
Here, also for short, $\bbox{\hat{I}}^{v2}\equiv I^v(I^v+1)$. The
dimensional $C$-constants are expressed {\em via} non-dimensional $R$%
-constants according to Eq.~(\ref{Rform}). $\hbar S_{\varsigma \varsigma
^{\prime }}=\mu _I^\varsigma \mu _I^{\varsigma ^{\prime }}/r_{\varsigma
\varsigma ^{\prime }}^3$ and $\mu _I^\varsigma $ is defined before (\ref{Vsr}%
). The subscript $O$ designates the intersection point of molecular symmetry
axis with the base resting on three hydrogen nuclei. {$\Delta J=J^{\prime }-J
$} and {$\Delta K=K_r^{\prime }-K_0=|\Delta K|(-1)^{r+\Delta K}$} by $%
|\Delta K|=1,2$. It is useful to have in mind that, in the resonance
conditions $\omega _{J^{\prime }K^{\prime }/JK}\simeq 0$,
\begin{equation}
\Delta J/\Delta K\simeq -B_K(K^{\prime }+K)/B_J(J^{\prime }+J+1).
\label{res_cond}
\end{equation}
In particular, one can see that $\Delta J$ and $\Delta K$ have always
different signs for fluoromethane (as prolate top). Formally, the
symmetrization changes only one factor to $C_{(-)\underline{z}}^{{\rm H}E}$
in expression (92) from \cite{Gus95}, i.e.,
\[
\left\{
\begin{array}{ccc}
J^{\prime } & 1 & J \\
1 & J & 1
\end{array}
\right\} \sqrt{\bbox{\hat{J}}^2[J]}
\]
is replaced by
\begin{equation}
\frac 12\left[ \left\{
\begin{array}{ccc}
J^{\prime } & 1 & J \\
1 & J & 1
\end{array}
\right\} \sqrt{\bbox{\hat{J}}^2[J]}-\left\{
\begin{array}{ccc}
J & 1 & J^{\prime } \\
1 & J^{\prime } & 1
\end{array}
\right\} \sqrt{\bbox{\hat{J}}^{\prime 2}[J^{\prime }]}\right] =\frac{\delta
_{|\Delta J|,1}}2\frac{\Delta J(J^{\prime }+J+1)}{\sqrt{6}}.
\end{equation}
The analogical factor to $C_{\underline{\bot }}^{{\rm H}E}$ and $C_{(+)%
\underline{z}}^{{\rm H}E}$, in explicit form, is
\begin{equation}
\left\{
\begin{array}{ccc}
J^{\prime } & 1 & J \\
1 & J & 2
\end{array}
\right\} \sqrt{\bbox{\hat{J}}^2[J]}=-\frac{\delta _{|\Delta J|,1}}2\sqrt{%
\frac{(J^{\prime }+J+3)(J^{\prime }+J-1)}{10}}
\end{equation}
and not changed. Now, taking into account also the definite symmetry of
Wigner coefficients, i.e.,\footnote{%
It is sometimes convenient for the negative sign to be denoted as overbar.}
\[
\langle J^{\prime }K^{\prime }|\varkappa \Delta K~JK\rangle \sqrt{[J]}%
(-1)^{J+K}=\langle JK|\varkappa \overline{\Delta K}~J^{\prime }K^{\prime
}\rangle \sqrt{[J^{\prime }]}(-1)^{J^{\prime }+K^{\prime }},
\]
one can see the expression (\ref{conv}) is invariant under the transposition
$(J^{\prime },K^{\prime }\leftrightarrow J,K)$.

\section{Conversional spin-rotation constants of fluoromethane}

It is convenient to collect the quantitative characteristics of the most
important accidental resonances for conversional transitions of
fluoromethane in Table~\ref{tconv}. The contribution of the individual
resonance reaches maximum for pressure
\begin{equation}
P_{J^{\prime }K^{\prime }/JK}^{(\text{max})}=\omega _{J^{\prime }K^{\prime
}/JK}/\Gamma _{J^{\prime }K^{\prime }/JK}^{(P)}.
\end{equation}
All the resonances have $\Gamma _{J^{\prime }K^{\prime }/JK}\ll |\omega
_{J^{\prime }K^{\prime }/JK}|$ for pressure $\displaystyle          P\ll
\min_{J^{\prime }K^{\prime }/JK}P_{J^{\prime }K^{\prime }/JK}^{(\text{max})}$%
, i.e. much less than $17~{\rm Torr}$ for ${}^{12}{\rm CH}_3{\rm F}$ and $7~%
{\rm Torr}$ for ${}^{13}{\rm CH}_3{\rm F}$. $K$-doubling in the parity for
mentioned resonances with $K=3$ does not matter practically, i.e., $\omega
_{J_pK}\simeq \omega _{JK}$ for all $K$. Thus, using (\ref{parity}), one can
sum over $p$ in (\ref{gamma}), i.e.,
\begin{equation}
\sum_p(1+\delta _{K,0})\{\ldots \}=2\{\ldots \}.
\end{equation}
Let us write out the explicit dependence of the $P$-normalized conversional
rate on spin-rotation $R$-constants:
\[
\gamma ^{(P)}=\sum_{J^{\prime }K^{\prime }/JK}(\gamma _{\text{sr}(J^{\prime
}K^{\prime }/JK)}^{(P)}+\gamma _{\text{ss}(J^{\prime }K^{\prime
}/JK)}^{(P)})
\]
with
\begin{equation}
\gamma _{\text{sr}(J^{\prime }K^{\prime }/JK)}^{(P)}=a_{\underline{\bot }%
(J^{\prime }K^{\prime }/JK)}^{{\rm H}R}(R_{\underline{\bot }}^{{\rm H}%
E})^2+\sum_{\sigma ^{\prime }\sigma }a_{\underline{z}(J^{\prime }K^{\prime
}/JK)}^{{\rm H}R(\sigma ^{\prime }\sigma )}R_{(\sigma ^{\prime })\underline{z%
}}^{{\rm H}E}R_{(\sigma )\underline{z}}^{{\rm H}E}.  \label{quadR}
\end{equation}
Here the quadratic form are positive semi-definite, i.e. $\geq 0$. By $%
\Gamma _{J^{\prime }K^{\prime }/JK}\ll |\omega _{J^{\prime }K^{\prime }/JK}|$%
, the coefficients $a_{\underline{\bot }(J^{\prime }K^{\prime }/JK)}^{{\rm H}%
R}$ and $a_{\underline{z}(J^{\prime }K^{\prime }/JK)}^{{\rm H}R(\sigma
^{\prime }\sigma )}$, as well as the conversional rate terms $\gamma _{\text{%
sr}(J^{\prime }K^{\prime }/JK)}^{(P)}$ and $\gamma _{\text{ss}(J^{\prime
}K^{\prime }/JK)}^{(P)}$, are constant and given in Table~\ref
{tconv:continue} for every resonance separately.

For ${\rm CH}_3{\rm F}$ one can calculate
\begin{equation}
S_{{\rm HF}}=13.4~{\rm kHz},\quad S_{{\rm HC}}=22.2~{\rm kHz},\quad S_{{\rm %
HH}}=19.9~{\rm kHz}.  \label{SS}
\end{equation}
Hence, for isotope ${}^{13}{\rm CH}_3{\rm F}$, it follows
\begin{equation}
^{13}\gamma _{\text{ss}}^{(P)}=1210~\mu {\rm Hz}/{\rm Torr},
\end{equation}
i.e. $62\%$ of (\ref{13exp}). For the rest, we obtain an equation
\begin{equation}
^{13}\gamma _{\text{sr}}^{(P)}=12.5(R_{\underline{\bot }}^{{\rm H}%
E})^2\times 10^{18}~\mu {\rm Hz}/{\rm Torr}=732~\mu {\rm Hz}/{\rm Torr}.
\end{equation}
Its solution is
\begin{equation}
\left| R_{\underline{\bot }}^{{\rm H}E}\right| \simeq 7.65~{\rm ppb}.
\label{Rbot}
\end{equation}
Similarly, for isotope $^{12}{\rm CH}_3{\rm F}$, we have
\begin{equation}
^{12}\gamma _{{\rm ss}}^{(P)}=2.85~\mu {\rm Hz}/{\rm Torr},
\end{equation}
i.e. $8\%$ of (\ref{12exp}). For the rest, we obtain another equation
\begin{eqnarray}
^{12}\gamma _{\text{sr}}^{(P)} &=&\left[ 0.151(R_{\underline{\bot }}^{{\rm H}%
E})^2+0.92(R_{(+)\underline{z}}^{{\rm H}E})^2\right.  \nonumber \\
&&\left. -1.87\times 10^{-3} R_{(+)\underline{z}}^{{\rm H}E}R_{(-)\underline{%
z}}^{{\rm H}E}+10^{-6}(R_{(-)\underline{z}}^{{\rm H}E})^2\right] \times
10^{18}~\mu {\rm Hz}/{\rm Torr}=32.45~\mu {\rm Hz}/{\rm Torr}.
\end{eqnarray}
Using (\ref{Rbot}) and setting $R_{(-)\underline{z}}^{{\rm H}E}=0$, we come
to the solution
\begin{equation}
|R_{(+)\underline{z}}^{{\rm H}E}|\simeq 5.07~{\rm ppb}.
\end{equation}

\section{Comparison of spin-rotation constants of fluoro\-meth\-ane and
(deutero)\-meth\-ane}

Here the conversional spin-rotation constants of fluoromethane are found for
the first time. To compare them with known respective constants of methane
(or deuteromethane), we must equally orientate both molecules so that F of $%
{\rm CH}_3{\rm F}$ corresponds to ${\rm H}^4$ of ${\rm CH}_3{\rm H}$ (or D
of ${\rm CH}_3{\rm D}$). The spin-rotation constants of ${\rm CH}_4$ are
usually cited in the following form \cite{YOR71}:
\[
C^{{\rm H}A_1}=g_I^{{\rm H}}\tilde{B}(2R_{\underline{\bot }}^{{\rm H}}+R_{%
\underline{z}}^{{\rm H}})/3=10.4(1)~{\rm kHz},\quad C^{{\rm H}F}=g_I^{{\rm H}%
}\tilde{B}(R_{\underline{\bot }}^{{\rm H}}-R_{\underline{z}}^{{\rm H}%
})=18.5(5)~{\rm kHz}.
\]
From here, using $\tilde{B}=314.2(1)~{\rm GHz}$ (see, e.g., \cite{CP80}),
one can obtain
\begin{equation}
\underline{\pmb{\sf R}}^{{\rm H}^4}(\underline{z}_{{\rm H}^4})=\left[
\begin{array}{cc}
R_{\underline{\bot }}^{{\rm H}}\hat{\pmb{\sf 1}}_2 & \bbox{0} \\
\bbox{0}^{\top } & R_{\underline{z}}^{{\rm H}}
\end{array}
\right] =\left[
\begin{array}{cc}
9.46(11) \hat{\pmb{\sf 1}}_2 & \bbox{0} \\
\bbox{0}^{\top } & -1.1(2)
\end{array}
\right] ~{\rm ppb}  \label{H4:CH4}
\end{equation}
and
\begin{eqnarray}
\underline{\pmb{\sf R}}^{{\rm H}^1}(\underline{z}_{{\rm H}^4}) &=&\frac 19%
\left[
\begin{array}{cc}
(5R_{\underline{\bot }}^{{\rm H}} +4R_{\underline{z}}^{{\rm H}}) \hat{%
\pmb{\sf 1}}_2-4(R_{\underline{\bot }}^{{\rm H}}-R_{\underline{z}}^{{\rm H}})%
\bbox{\hat{\sigma}}_{{\rm v}}^{{\rm H}^1} & \sqrt{8}(R_{\underline{\bot }}^{%
{\rm H}}-R_{\underline{z}}^{{\rm H}})\bbox{n}^{{\rm H}^1} \\
\sqrt{8}(R_{\underline{\bot }}^{{\rm H}}-R_{\underline{z}}^{{\rm H}})\bbox{n}%
^{1\top } & (8R_{\underline{\bot }}^{{\rm H}}+R_{\underline{z}}^{{\rm H}})
\end{array}
\right]  \nonumber \\
&=&\left[
\begin{array}{cc}
4.76(11) \hat{\pmb{\sf 1}}_2-4.69(13)\bbox{\hat{\sigma}}_{{\rm v}}^{{\rm H}%
^1} & 3.31(9)\bbox{n}^{{\rm H}^1} \\
3.31(9)\bbox{n}^{{\rm H}^1\top } & 8.27(10)
\end{array}
\right] ~{\rm ppb}  \label{H1:CH4}
\end{eqnarray}
where $\bbox{\hat{\sigma}}_{{\rm v}}^{{\rm H}^1}$ and $\bbox{n}^{{\rm H}^1}$
are defined in (\ref{RH1}). One may see $R_{\underline{xx}}^{{\rm H}^1}\ll
R_{\underline{yy}}^{{\rm H}^1}$. Take notice $R_{\underline{\bot }}^{{\rm H}%
E}={}_{(\text{n})}R_{\underline{\bot }}^{{\rm H}E}+{}_{(\text{e})}R_{%
\underline{\bot }}^{{\rm H}E}<0$ as well as nuclear term ${}_{(\text{n})}R_{%
\underline{\bot }}^{{\rm H}E}$ easily calculated from \cite{GMTVV54}. To
make complete the set of methane constants, we cite from \cite{GKRS97}
\begin{equation}
\underline{\pmb{\sf R}}^{{\rm C}}(\underline{z}_{{\rm H}^4})=-28(7) \hat{%
\pmb{\sf 1}}_3~{\rm ppb}.  \label{C:CH4}
\end{equation}
The diagonal components of tensor $\underline{\pmb{\sf C}}^\varsigma $ for $%
{\rm CH}_3{\rm F}$ can be found in \cite{WMK71}. Using the components of $%
\widetilde{\pmb{\sf B}}$ from (\ref{tildeB}), we obtain
\begin{equation}
\underline{\pmb{\sf R}}^{{\rm F}}(\underline{z}_{{\rm F}})=\left[
\begin{array}{cc}
15(7) \hat{\pmb{\sf 1}}_2 & \bbox{0} \\
\bbox{0}^{\top } & -31.4(8)
\end{array}
\right] ~{\rm ppb},  \label{F:CH3F}
\end{equation}
\begin{equation}
\underline{\pmb{\sf R}}^{{\rm H}^1}(\underline{z}_{{\rm F}})\simeq \left[
\begin{array}{cc}
2.9(54) \hat{\pmb{\sf 1}}_2-7.65\bbox{\hat{\sigma}}_{{\rm v}}^{{\rm H}^1} &
5.07\bbox{n}^{{\rm H}^1} \\
5.07\bbox{n}^{{\rm H}^1\top } & 8.4(4)
\end{array}
\right] ~{\rm ppb}.  \label{H:CH3F}
\end{equation}
Here we have added the above obtained off-diagonal components. Their signs
are supposed to be just like the ones of (deutero)methane constants in (\ref
{H1:CH4}). To make complete also the set of fluoromethane constants, we can
use (\ref{shielding}) and data from \cite{AD74,Fly78}. As a result, one may
estimate
\begin{equation}
\underline{\pmb{\sf R}}^{{\rm C}}(\underline{z}_{{\rm F}})\simeq \left[
\begin{array}{cc}
-80\times \hat{\pmb{\sf 1}}_2 & \bbox{0} \\
\bbox{0}^{\top } & -35
\end{array}
\right] ~{\rm ppb}.  \label{C:CH3F}
\end{equation}

\section{Conclusion}

Summing up, we notice the following. Unlike \cite{Gus95}, the symmetrized
spin-rotation contribution to nuclear spin-modification conversion in ${\rm C%
}_{3{\rm v}}$-symmetry molecules has been produced. But the symmetrization
is important only for low $J$.

Both conversional spin-rotation constants $R_{\underline{\bot }}^{{\rm H}E}$
and $R_{(+)\underline{z}}^{{\rm H}E}$ obtained here for fluoro\-meth\-ane
are roughly one and half times more than the respective constants of
(deu\-tero)\-meth\-ane. This difference in reality is slightly less even
because to estimate the conversional rate we have taken into account only
the most sharp accidental resonances, i.e. the most coupled rotational
levels of nuclear spin modifications. There is a more complete set\footnote{%
See also the footnote on p.~\pageref{FK}.} of these resonances with $J\leq
80 $ in \cite{Nag98}.

The degree of asymmetry for spin-rotation tensor $\pmb{\sf R}^h$ still
remains unclear. The asymmetry of $\pmb{\sf R}^h$ is absent for methane
because of sufficiently high ${\rm T}_{{\rm d}}$-symmetry. But we have only
supposed the antisymmetric term $\pmb{\sf R}^h_{(-)}$ to be zero for
fluoromethane.

The constant $R^{{\rm H}E}_{\underline\bot}$ in contrast to both $R^{{\rm H}%
E}_{(\pm)\underline z}$ manifests itself in spectra as hyperfine doubling of
rotational levels with $K=1$ \cite{GMWS:54,TS55}. One can find it, e.g.,
using the magnetofield spectrum of nonlinear-optical resonance \cite{Gus95}.
More promising alternative way of finding all the three conversional
spin-rotation constants was described by authors of \cite{BIAIC98}. They had
suggested to measure the nuclear spin conversion of fluoromethane in the
presence of electric field. The Stark-induced crossings of rotational levels
would allow to disentangle the conversional spin-rotation and spin-spin
contributions. The spin-spin contribution can be used as a scale to measure
the spin-rotation one in this method. The last circumstance is especially
attractive since it allows to remove the problems connected with the
broadening of accidental resonances.

\acknowledgments
The author is indebted to P.~L. Chapovsky and L.~V. Il'ichov for the useful
remarks and assistance. This work was financially supported by the Russian
Fund for Basic Researches (Grant No. 98-03-33124a).

\newpage

\newpage


\begin{table}[tbp]
\caption[The most weighty conversional resonances.]{The most weighty
conversional resonances for ${}^{12,13}{\rm CH}_3{\rm F}$ isotopes.}
\label{tconv}%
\begin{tabular}{ccccddc}
Isotope & $J',K'/J,K$~\tablenotemark[1] &
$\Delta J/\Delta K$ &
$\omega_{J'K'}/\omega_{JK}$~\tablenotemark[2] &
$\omega_{J'K'/JK}$~\tablenotemark[2] &
$\Gamma^{(P)}_{J'K'/JK}$~\tablenotemark[3] &
$P^{(\text{max})}_{J'K'/JK}$~\tablenotemark[4] \\ \tableline
$^{12}{\rm CH}_3{\rm F}$ & $9,2/10,0$ & $\overline1/2$
& $2\,816.8/2\,808.2$ & $8$.$592$ & $17$.$4$ & 494\\
--- & $15,7/17,6$ & $\overline2/1$ & $12\,476/12\,474$
& $1$.$756$ & $18$.$8$ &  $93$  \\
--- & $28,5/27,6$ & $1/\overline1$ & $23\,931/23\,930$
& $1$.$185$ & $8$.$8$  & $135$ \\
--- & $51,4/50,6$ & $1/\overline2$ & $69\,356/69\,356$
& $-0$.$048$ & $2$.$8$     &  $17$ \\[2mm]
${}^{13}{\rm CH}_3{\rm F}$ & $11,1/9,3$ & $2/\overline2$
& $3\,411.3/3\,411.2$  & $0$.$130$  & $17$.$4$ & $7$ \\
--- & $21,1/20,3$ & $1/\overline2$ & $11\,605/11\,605$
& $-0$.$352$ & $16$. & $22$
\end{tabular}
\tablenotetext[1]{In accordance with \cite{Cha91}.} \tablenotetext[2]{$({\rm %
GHz})$.} \tablenotetext[3]{$({\rm MHz}/{\rm Torr})$. See \cite{TEB79}.} %
\tablenotetext[4]{$({\rm Torr})$.}
\end{table}

\begin{table}[tbp]
\caption[The coefficients and conversional contributions.]{The coefficients $%
a^{{\rm H}R}_{\protect\underline{\bot}(J^{\prime}K^{\prime}/JK)}$ and $a^{%
{\rm H}R(\sigma^{\prime}\sigma)}_{\protect\underline{z}(J^{\prime}K^{%
\prime}/JK)}$ of quadratic $R$-forms in (\ref{quadR}). The conversional
spin-rotation and spin-spin contributions.\tablenotemark[1]}
\label{tconv:continue}
\begin{tabular}{cccccc}
$a^{{\rm H}R}_{\underline{\bot}(J^{\prime}K^{\prime}/JK)}$~\tablenotemark[2]
& $a^{{\rm H}R(++)}_{\underline{z}(J^{\prime}K^{\prime}/JK)}$~%
\tablenotemark[2] & $2a^{{\rm H}R(+-)}_{\underline{z}(J^{\prime}K^{%
\prime}/JK)}$~\tablenotemark[2] & $a^{{\rm H}R(--)}_{\underline{z}%
(J^{\prime}K^{\prime}/JK)}$~\tablenotemark[2] & $\gamma^{(P)}_{\text{sr}%
(J^{\prime}K^{\prime}/JK)}$~\tablenotemark[3] & $\gamma^{(P)}_{\text{ss}%
(J^{\prime}K^{\prime}/JK)}$~\tablenotemark[3] \\
\tableline $0.011$ & $0$ & $0$ & $0$ & $0.64$ & $0.91$ \\
$0$ & $0$ & $0$ & $0$ & $0$ & $0.71$ \\
$0$ & $0.92$ & $-1.87\times10^{-3}$ & $10^{-6}$ & $23.65$ & $0.78$ \\
$0.14$ & $0$ & $0$ & $0$ & $8.19$ & $0.45$ \\[2mm]
$0$ & $0$ & $0$ & $0$ & $0$ & $800$ \\
$12.5$ & $0$ & $0$ & $0$ & $732$ & $410$%
\end{tabular}
\tablenotetext[1]{The rows of Table~\ref{tconv} are continued here.} %
\tablenotetext[2]{$(10^{18}\mu{\rm Hz}/{\rm Torr})$.} \tablenotetext[3]{$(\mu%
{\rm Hz}/{\rm Torr})$.}
\end{table}

\end{document}